\documentclass[12pt,preprint]{aastex}
\usepackage{graphicx}
\usepackage{epstopdf}
\newcommand{\heir}{He I~$\lambda 10830$}
\newcommand{\heopt}{He I~$\lambda 5876$}

\newcommand{\kms}{km~s$^{-1}$}
\newcommand{\etal}{{\it et al.}}

\begin{document}
\slugcomment{to appear in AIP Proceedings of Cool Stars, Stellar Systems and the Sun \# 15}
\title {Winds and Accretion in Young Stars}


\author{Suzan Edwards \altaffilmark{1}}
\altaffiltext{1}{Astronomy Department, Smith College, Northampton, MA 10163, sedwards@smith.edu}

\begin{abstract}
Establishing the origin of accretion powered winds from forming stars is critical for understanding angular momentum evolution in the star-disk interaction region. Here, the high velocity component of accretion powered winds is launched and accreting stars are spun down, in defiance of the expected spin-up during magnetospheric accretion.  T Tauri stars in the final stage of disk accretion offer a unique opportunity to study the connection between accretion and winds and their relation to stellar spindown. Although spectroscopic indicators of high velocity T Tauri winds have been known for decades, the line of He I 10830 offers a promising new diagnostic to probe the magnetically controlled star-disk interaction and wind-launching region. The high opacity and resonance scattering properties of this line offer a powerful probe of the geometry of both the funnel flow and the inner wind that, together with other atomic and molecular spectral lines covering a wide range of excitation and ionization states, suggests that the magnetic interaction between the star and disk, and the subsequent launching of the inner high velocity wind, is sensitive to the disk accretion rate.

\end{abstract}

\maketitle


\section{Introduction  }

The early evolution of a young star is characterized by active accretion from a proto-planetary disk and ejection of collimated mass outflows. A robust correlation between indicators of mass accretion and outflow  reveals that outflows are accretion-powered  although the means by which accretion energy is tapped to launch the wind remains a mystery\cite{jf06}.  A fundamental question is whether accretion powered outflows transport angular momentum from the disk, allowing it to evolve as an accretion disk, or whether they are nature's means of keeping the accreting star from rotating near break-up. Accretion powered outflows  also may  be agents in heating the disk atmosphere \cite{glass04}, in hastening disk dissipation \cite{h00}, and in disrupting infalling material from the collapsing molecular cloud core \cite{t84}.

While observations of the full mass range of young stars show correlations between diagnostics of outflow and accretion, the underpinning of the accretion-outflow connection is provided by the young low mass T Tauri stars (TTS) with small enough extinctions to be optically visible. Among these stars accretion luminosities and corresponding mass accretion rates are derived from the optical/ultraviolet emission in excess of photospheric radiation (``veiling'') that is attributed to accretion shocks at the terminus of magnetospheric footpoints linking the disk and the star \cite{bouv07}. Mass ejection rates from accreting TTS are most reliably determined from their blueshifted forbidden lines which often show a high velocity component arising from shocked gas in a  ``microjet''.  Although even for well studied TTS in a restricted range of mass and evolutionary state the ejection/accretion ratio, ${\dot M}_{eject}/{\dot M}_{acc}$, is not well constrained, the best estimate to date suggests a ratio on the order of  0.1, based on the survey of Hartigan et al. (1995) but with mass accretion rates revised upward following Gullbring and collaborators\cite{gul98,cabrit07}. The fundamental nature of the accretion-outflow connection is further illustrated by the weak emission TTS, with optically thin or absent inner disks, which also lack both detectable veiling and forbidden line emission, demonstrating that when there is no accretion of material onto the stellar surface, there is also no outflow.

Most theoretical models for the origin of accretion powered winds are motivated by the need to transport angular momentum away from either the disk or the star. There is no doubt that {\it both} must be removing angular momentum during the accretion phase, but a key question is what role do outflows play in these processes? In order for a disk to accrete, it must transport angular momentum outwards. A magneto-centrifugal wind arising over a wide range of disk radii is an elegant way to do to this, but other processes for enabling disk accretion are also compelling, such as MRI or density waves \cite{pud07}. Similarly, in order for an accreting star to have a slow spin that is both an order of magnitude below break-up and slower than its brethren that have already dissipated their accretion disks, angular momentum must be removed (or deflected) from the star while it is accreting\cite{reb06}. A magneto-centrifugal disk wind arising over a narrow range of radii near the corotation radius, using a magnetic field hijacked from the star but now opening from the ``X point'' offers an elegant way to solve the stellar spin-down problem \cite{shu00}. On the other hand, an accreting star might instead shed angular momentum via an accretion-powered stellar wind following magnetic field lines emerging radially from the star \cite{matt05}, or via coupling of the stellar magnetosphere to the disk at radii both inside and outside the co-rotation radius \cite{cc93}.

Of course some combination of these processes may occur. However, the ubiquity of the accretion-outflow connection signifies that outflows are an integral component of the angular momentum evolution of an accreting star-disk system and it is thus imperative to establish whether they originate from the star or the disk.  In this short review, I will focus on observational probes that speak to the origin of accretion powered winds by looking at spectroscopic diagnostics arising from the inner AU of a star-disk system. Since each genre of wind model has distinctive launch sites, it may be possible to identify which is operational by assessing whether winds arise (1) from the disk over a range of radii extending from the inner truncation radius to a few AU, (2) from the disk over a narrow annulus near corotation, or (3) from the star.  The ultimate goal is to determine the role of winds in the spindown of an accreting star and in enabling disk accretion.

\section{Observational Clues to Outflow Origin}

Observations of the collimation and kinematic structure of microjets within 10-100 AU of the star are one means of probing wind origins \cite{ray}.  Although their spatially resolved kinematic structure has been attributed to origin in extended disk winds \cite{cof07}, alternate interpretations of the poloidal and toroidal velocities are also viable \cite{cerq06, cab06,tak07}.  Moreover, if extended disk winds are the dominant source of outflowing gas in the microjets, it is surprising that the numerous ro-vibrational molecular emission lines, formed in the warm disk chromosphere over radii from  0.1 to 2 AU,  show no sign of outflow \cite{naj07}. In contrast, there is abundant spectroscopic evidence for high velocity winds originating in the vicinity of the star-disk interaction region where the stellar magnetic field is thought to truncate the disk and channel accreting material to the star via funnel flows. All of the potential sources for winds are expected to operate in this region, but, if present, will have distinctive outflow geometries.

Some wind diagnostics from this region, i.e.  blueshifted absorption in strong permitted optical and NUV lines such as H$\alpha$, CaII H\&K, MgII h\&k and NaD have been recognized as mass loss indicators for decades \cite{cal97,ard02}, while others, such as \heir, are relatively new probes \cite{edw03,dup05}. Progress interpreting  lines from the star-disk interaction region has been thwarted by the almost bewildering array of profile morphologies due to hybrid formation in the funnel flow, accretion shock, and wind, viewed at a variety of inclinations.  Another contributor to profile diversity is the factor of 100-1000 in accretion rates characterizing the T Tauri class, where both the strength and morphology of emission lines scale roughly with the veiling/mass accretion rate \cite{ab00,edw06}. Temporal variations are also a factor, arising both from rotating non-axisymmetric magnetic structures and small variations in the disk accretion rate, although the magnitude of the variations displayed by an individual star is typically small compared to differences from extremes in disk accretion rates among the T Tauri class. Synoptic studies in a few low accretion rate stars suggest simple phased relations between infalling and outflowing gas, however among stars with high disk accretion rates synoptic data show chaotic behavior \cite{jk97}. 

\begin{figure}
  \includegraphics[height=5in,width=3in,angle=90]{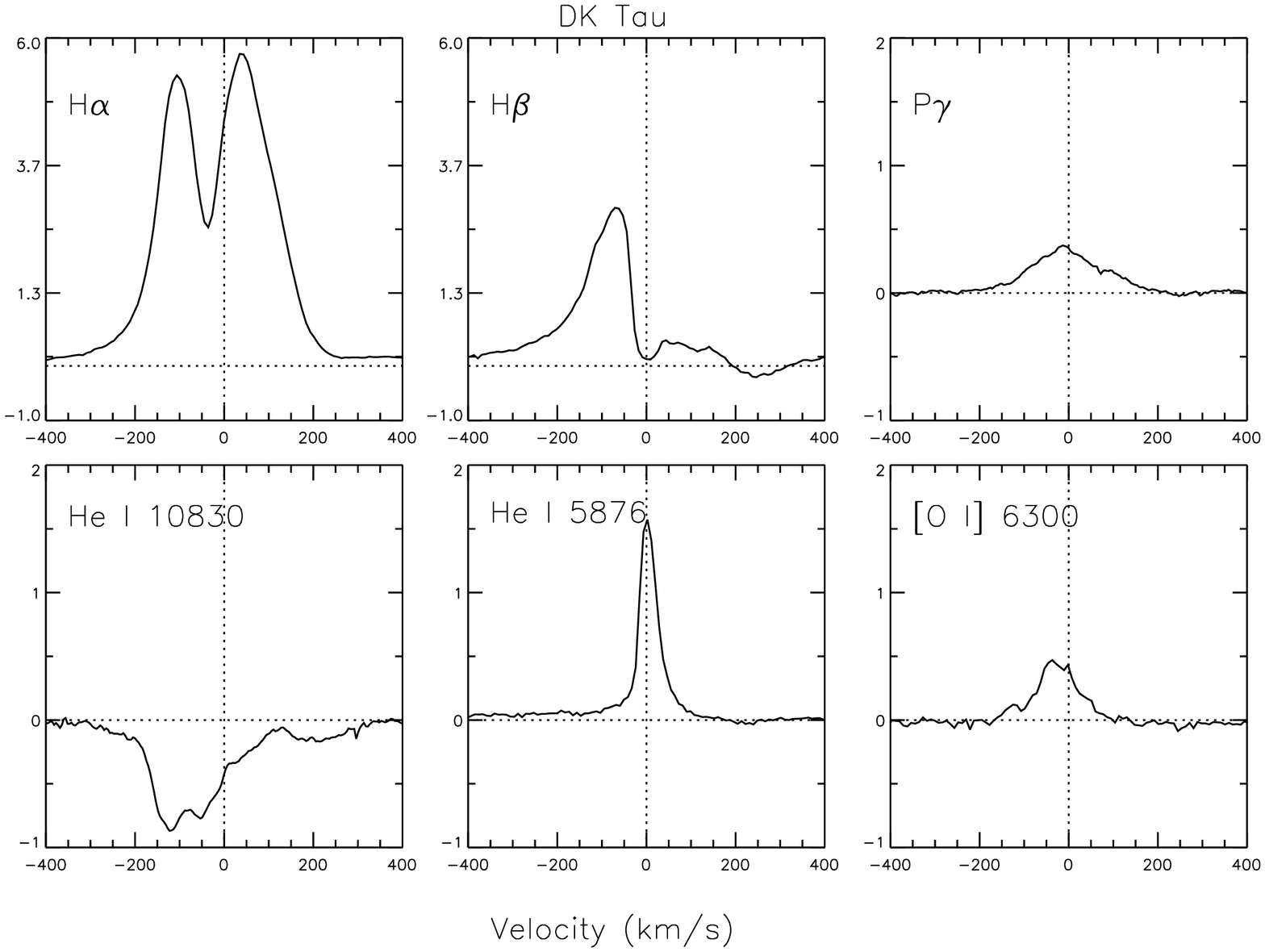}
  \caption{Simultaneous profiles of accreting TTS DK Tau showing evidence for winds, magnetospheric accretion, and accretion shocks (see text). Velocities are relative to the stellar photosphere and intensities are residuals above the photosphere.}
\end{figure}

An illustration of the variety of profile morphologies among different spectral features in one star is shown in Figure 1 for the moderate accretion rate TTS DK Tau, acquired simultaneously with the Keck I and Keck II  telescopes. The blueshifted forbidden emission line of [O I] 6300 arises at least partially in a microjet, tracing outflowing gas at distances in excess of 10 AU. The hydrogen profiles in the upper panel,  H$\alpha$, H$\beta$, and P$\gamma$, show prominent broad emission that is typically attributed to infalling gas in the funnel flow \cite{muz01}, although additional regions may also contribute \cite{bek,whelan04,edw06}. The H$\alpha$ profile also shows a narrow blue absorption cutting into the emission suggestive of a wind, but a more definitive wind indicator is the broad, deep subcontinuum blue absorption seen at \heir. Red absorption from the funnel flow is seen both in H$\beta$ and \heir, while \heopt\ is dominated by narrow emission that is slightly redshifted, indicating an origin in the post shock region at the footpoint of the funnel flow.

A promising approach to study inner high velocity winds is FUV lines, as shown by a few studies targeted at individual stars. The high accretion rate TTS RU Lup \cite{herczeg05} has numerous neutral and singly ionized metallic lines with a clear P Cygni character, with blue absorption velocities from -70 to -200 \kms, in some cases penetrating the continuum. The low accretion rate TTS TW Hya shows P Cygni profiles in lines of higher excitation, up to C II and possibly O VI, indicating outflowing gas with temperatures of at least 30,000 K \cite{dup05,jk07}. In addition to P Cygni profiles, blueshifted emission has been attributed to winds in a few stars, including O VI \cite{gun08} and semi-forbidden lines of SiIII], C III], and [O II]  \cite{gomez07}. Since to date the FUV studies have targeted only a small number of stars, it is too early to make conclusions regarding their role in accretion-powered outflows. Judging from the well studied optical lines, where diverse profile morphologies are found among different lines in a single star and between stars of different accretion rates, it is clear that  progress in interpreting FUV lines from the star-disk interaction region requires working with a sizable sample of accreting stars so that effects of viewing angle and disk accretion rate can be sorted out.

In sum, although there is clear evidence for high velocity winds emerging from the star-disk interaction zone the complexity of most of the strong permitted optical and NUV lines makes it difficult to extract quantitative information on the inner wind. Order of magnitude estimates for mass loss rates have been made for those stars with blueshifted absorption at H$\alpha$, with ${\dot M}_{wind}$ $\sim 10^{-9}$ $M_{\odot}$~yr$^{-1}$ for stars with average mass accretion rates of ${\dot M}_{acc}$ $\sim 10^{-8}$ $M_{\odot}$~yr$^{-1}$, although ${\dot M}_{wind}$ is as high as $10^{-7}$ $M_{\odot}$~yr$^{-1}$ for some high accretion rate stars \cite{cal97,kur06}. The hybrid nature of these lines makes mass loss estimates uncertain and identifying the wind launch region impossible. Looking to the future, the less well studied ultraviolet appears to offer a rich variety of probes of outflowing gas when large enough surveys can be done.  At present, however, helium lines offer the best means to date for elucidating the origin of high velocity inner winds, as discussed in the following section.

\subsubsection{Tracing the Inner Wind with Helium}

A new diagnostic of the inner wind region in accreting stars, \heir, has the potential to break through the ambiguity in determining where winds are launched in the star-disk interaction region.  A NIRSPEC survey at 1$\mu$m of 38 accreting T Tauri stars spanning a wide range of disk accretion rates shows that \heir\ has a far higher incidence of P Cygni profiles than any other line observed to date \cite{edw06}, often with a velocity structure reminiscent of spherical stellar wind models developed over a decade ago that failed to explain H$\alpha$ profiles in TTS \cite{hart90}. The \heir\ line offers a unique probe of the geometry of the inner wind because it is formed under conditions resembling resonance scattering, with a high opacity in its metastable lower level ($2s^3S$)  and only one allowed radiative exit from the upper level. Thus it readily forms an absorption feature via  scattering of the 1$\mu$m continuum under most conditions. Any additional source of emission is very modest in comparison to $H\alpha$, a line with net emission equivalent widths 10-100 times greater than \heir\ \cite{edw03}. A further advantage offered by \heir\ is that it is restricted to  a more limited formation region than H$\alpha$, one either of high excitation or close proximity to a source of ionizing radiation.

\begin{figure}
  \includegraphics[height=3in,width=4in,angle=0]{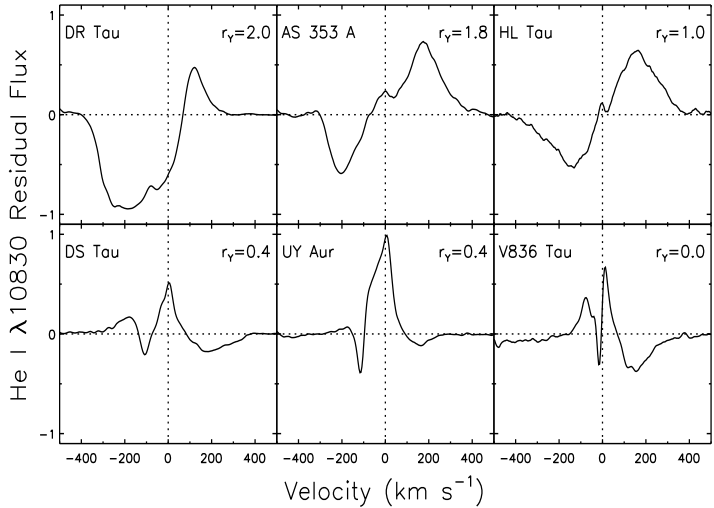}
  \caption{ Examples of \heir\ residual profiles and corresponding $1 \mu $ veiling, $r_Y$, for 6 accreting TTS. The upper row shows high veiling objects with P Cygni profiles characterized by deep and broad blue absorption and the lower row shows low veiling objects with narrow blue absorption. The latter also show red absorption from magnetospheric infall.}
\end{figure}

The $1\mu$ TTS NIRSPEC survey shows \heir\ to be an outstanding probe of both winds and infalling gas in the magnetosphere, as shown in Fig 2. Unlike other hybrid lines, the fact that this line is formed primarily by scattering makes it straightforward to separate the two contributions, which are both seen by absorption of the $1\mu$ continuum. Blueshifted absorption {\it below the continuum} is seen in 70 \% of the stars, in contrast to H$\alpha$ where the fraction is $\sim$10\%.  Subcontinuum redshifted absorption is seen in 50\% of the stars in contrast to 24\% in the neighboring $P\gamma$ line. The character of the blue absorptions is varied, in some stars it is remarkably broad and deep, with up to 90\% of the 1$\mu $ continuum absorbed over a velocity interval of 300-400 \kms, while others show narrow absorption with modest blueshifts.   There is no question that the wind probed by \heir\ derives from accretion, since it is not seen in non-accreting T Tauri stars and the strength of the combined absorption and emission correlates with the excess continuum veiling at $1 \mu$. The red absorptions also show a range of depths and breadths, with up to 60\% of the 
1$\mu $ continuum absorbed over a velocity interval of 200-300 \kms. There is also a veiling dependence on profile morphology.   While blueshifted absorption is seen among stars with both high and low 1$\mu $ veiling, narrow blue absorptions are primarily found among stars with low veiling. Similarly, red absorptions are almost absent among stars with high 1$\mu $ veiling (4\%) but common among stars with low veiling (66\%).

We have undertaken two studies that take advantage of the scattering properties of \heir\  to set constraints on the geometry of the inner wind and the magnetospheric accretion flows in accreting TTS. The first one, focussed on the wind geometry, models \heir\ profiles via monte carlo simulations including both pure scattering and in-situ emission for two inner wind geometries: (1) a disk wind emerging at a constant angle relative to the disk surface and (2) a stellar wind emerging radially away from the star \cite{kwan07}.  The second one calculates \heir\ profiles formed via pure scattering in funnel flows in dipolar geometries with a range of flow widths, impact latitudes, and filling factors \cite{fischer08}.

I first discuss the wind models, parameterized with a simple set of assumptions that allowed a variety of effects to be explored. For the disk wind, launching was confined between 2-6 R$_*$ since helium excitation is unlikely to persist at great distances, and angular velocities along streamlines maintained either rigid rotation or conserved angular momentum. The stellar wind was assumed to originate between 2-4 R$_*$, both spherical and polar geometries were examined, and the effects of disk shadowing were explored. The latter ranged from extreme (disk truncation radius inside the wind origination radius) to negligible (disk truncation radius $\ge$ 5 times the wind origination radius). In each case, profiles were computed under the assumption of pure scattering, where all the $1\mu$ continuum photons were assumed to come from the star, and also with the additional presence of in-situ emission.  

The resulting suite of wind profiles show morphologies resembling the variety found in the data.  As shown in Fig. 3, one distinguishing feature between profiles formed in stellar vs. disk winds is the breadth of the blue absorption.  Scattering in a disk wind results in narrow blue absorptions as any line of sight to the star  intercepts a limited interval of radial velocities in the wind while scattering in a stellar wind necessarily yields broad blue absorptions as each line of sight to the star intercepts the full acceleration region of the wind. The broad absorptions from the stellar wind will extend to the terminal velocity of the wind, while the centroids of the narrow disk wind absorptions can have either a large or small blueshift, depending on the view angle to the star and the opening angle of the wind.  Among the full sample of 38 accreting T Tauri stars there are roughly comparable numbers with blue absorptions at \heir\ resembling disk (30\%) or stellar (40\%) winds.

\begin{figure}
  \includegraphics[height=3in,width=3.5in,angle=0]{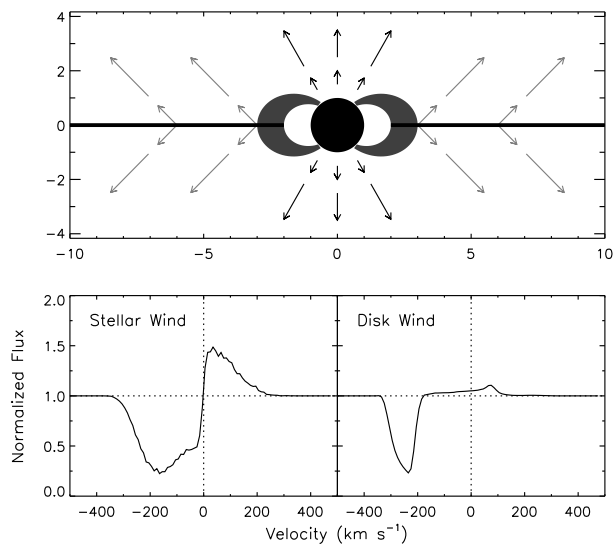}
  \caption{Illustration of the 2 genres of wind models explored to explain the morphology of \heir\ \cite{kwan07}. Classic P Cygni profiles result from pure scattering in a spherical stellar wind, while narrow blue absorption with little or no emission characterizes profiles formed by scattering in a disk wind.}
\end{figure}

A second distinguishing characteristic between stellar and disk winds is the morphology of the emission component above the continuum. For stellar winds a range of emission morphologies is possible, depending on whether the wind is dominated by scattering or in-situ emission (which can fill in the blue
absorption at low velocities), and whether disk shadowing is important (which will reduce the strength of the  emission feature). We see examples of all of these in the TTS survey. In contrast no stars show emission resembling a disk wind. This is not unexpected if the disk winds produce \heir\ only via scattering, as in that case any emission is spread out over such a range of velocities that it is extremely weak. However, in-situ emission from a disk wind is definitely ruled out, as it would either be entirely blueshifted or broad and double-peaked, depending on the view angle. Among the surveyed TTS, none show this behavior.  We conclude that in all cases showing narrow blue absorption from a disk wind the helium emission arises either from scattering in the funnel flow and/or scattering + emission from a polar stellar wind seen at sufficient inclination to have no line of sight absorption (bottom panel of Fig 2).

The finding that  \heir\ profiles suggest stellar winds in some TTS and disk winds in others would seem to be consistent with a picture where both are simultaneously present, as in Fig. 3, with the stellar wind confined to polar latitudes, magnetospheric funnel flows connecting the disk to the star at mid-latitudes, and a disk wind beginning at the disk truncation radius.  In this scenario, the character of the blue absorption would depend on view angle, with pole-on sources resembling stellar winds  and edge-on sources resembling disk winds, the latter accompanied by red absorptions. There are hints among the survey that such an aspect dependence is present, at least among some of the TTS with low $1\mu$ veiling. However, other factors must also be present, since the character of the observed profiles also depends on the veiling, which is a surrogate for the disk accretion rate. The low veiling sources more commonly show red absorptions and present the clearest cases for disk wind profiles (lower panel in Fig 2), while for high veiling sources red absorptions are rare and the clearest cases for stellar wind profiles, with classic P Cygni profiles formed entirely by scattering, are found (upper panel in Fig. 2). The means of launching such accretion powered stellar winds is not yet known, although the possibility that accretion energy generates MHD waves that can accelerate winds from stellar polar regions \cite{cran08} or that massive CME-like ejections are released as twisted magnetic loops, achieving coronal temperatures after partial draining of accreting gas in response to field inflation, violently break open \cite{hart08} are both intriguing new suggestions.

\subsubsection{Tracing Magnetospheric Accretion  with Helium}

\begin{figure}
  \includegraphics[height=2.4in,width=4in,angle=0]{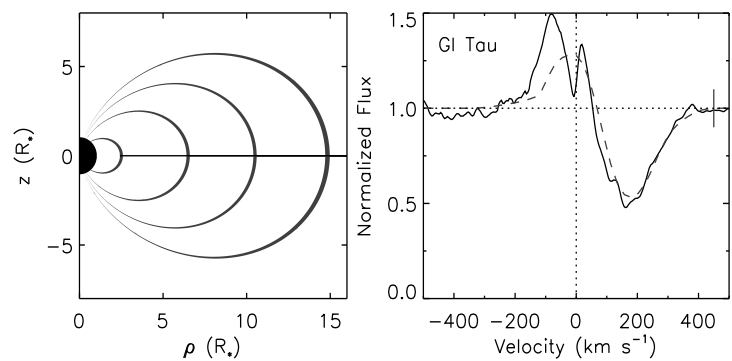}
  \caption{Illustration of magnetospheric infall in a wide dipolar flow whose volume is {\it dilutely} filled with accreting gas (on left).  Comparison of profiles for a dilute dipole (dotted) to a TTS  with deep and broad red absorption at \heir\ (solid, on right). Adapted from \cite{fischer08}.}
\end{figure}

In addition to being a unique diagnostic of the geometry of the inner wind in TTS,  \heir\ profiles are also effective probes of the accreting gas  via their redshifted absorptions, where continuum photons from the stellar photosphere are scattered by infalling gas in the funnel flow. The challenge for models is to account for the broad and deep red absorptions seen in many stars with low or absent $1\mu$ veiling. In the second modeling paper \cite{fischer08} profiles were calculated for lines formed via scattering by  gas in magnetospheric funnels under the assumption of dipolar flows, which have a well known relation between the radius in the disk intercepted by a field line and its corresponding impact latitude on the star. The effect of the width of the flow was explored, by varying the range of radii in the disk intercepted by mass loaded field lines and their corresponding magnetospheric filling factor $F$ on the star. Profiles were calculated both for cases where $F$ is equivalent to the hot shock filling factor $f$ on the stellar surface (i.e. the full volume of the funnel flow carries accreting gas terminating in hot shocks) and also for cases where $f$ $<$ $F$ (i.e. the volume of the funnel flow is only dilutely filled with accreting gas terminating in accretion shocks). 

In order to reproduce the observed broad and deep red absorptions at \heir\ wide flows were found to be required, with magnetospheric filling factors $F$ on the star of 10-20\%. However, the hot shock filling factors for the stars with the most extreme red absorptions are known to be much smaller than this, with $f \sim 1\%$ based on their optical veilings \cite{cal98}. The implication seems to be that dilutely filled magnetospheric flows are required to simultaneously explain the red absorptions and small optical veilings, in some cases with inner disk radii extending from 2$R_*$ out to the corotation radius with only 5-10\% of the volume filled with accreting gas, as illustrated in Fig. 4.

We interpret the need for wide but dilutely filled accretion flows as indicating that the funnel flows are comprised of numerous thin, widely spaced streamlets of infalling gas intercepting the star over a range of latitudes. If this is the case, it would provide a ready means of photoionizing all the accreting gas rather than just the skin of a thicker flow. In any case it is clear that the \heir\ profiles provide important new constraints for any model seeking to account for the geometry of magnetospheric accretion.

\section{Conclusions}

The high opacity and resonance scattering properties of \heir\ make it a valuable probe of the geometry of the inner high velocity wind and magnetospheric infall in accreting TTS.  The 3 main conclusions from the properties of this line are:
\begin{enumerate}
\item
Accretion powered stellar winds are present in many TTS although their contribution to the total mass ejection rate in jets and their role in stellar spindown is still uncertain. Their presence strengthens the argument of  Matt and Pudritz \cite{matt08} that they are required for spinning down accreting stars. If so, then the requisite ratio of ${\dot M}_{wind}/{\dot M}_{acc}$ $\sim ~0.1$ to achieve spin down would also mean accretion powered stellar winds are major players in the accretion/outflow connection.
\item
Magnetospheric accretion flows are very wide in some stars, but probably only dilutely fill the large flow volume. The implication of dilutely filled flows for stellar spindown via magnetic coupling to the disk both inside and outside co-rotation is an additional consideration for investigations of the efficacy of these models.
\item
The specifics of the star-disk interaction region appear to depend on the mass accretion rate. Helium lines in low accretion rate stars suggest co-existence of disk winds, polar stellar winds and funnel flows, consistent with the current paradigm for accreting stars. High accretion rate stars are less easily characterized, rarely showing  signatures for inner disk winds or magnetospheric accretion. This may relate to hints from numerical simulations that the character of accretion flows is sensitive to the accretion rate, which would likely have consequences for stellar spindown and wind launching \cite{rom08}. 
\end{enumerate}

The next step is to determine the physical conditions in the regions where helium emission and absorption are seen, with the goal of refining mass loss and mass accretion rates. Work in progress from our team (Hillenbrand, Fischer and Kwan) includes confronting simultaneous optical and near infrared helium and hydrogen line profiles with statistical equilibrium calculations for a 19-level helium atom and 6-level hydrogen atom to clarify whether mass loss rates from the inner wind traced by \heir\ are a major contributor to mass ejection rates inferred from the distant microjets and whether the stellar winds found in high accretion rate stars are sufficient to spin down stars during their accretion phase.


\vskip 0.2in
 {\bf Acknowledgements:} I gratefully acknowledge my
collaborators W. Fischer, L. Hillenbrand, and J. Kwan, without
whom the results on \heir\ would not exist, along with NASA OSS grant
NNG506GE47G  and the staff of the Keck Observatories.






\end{document}